\documentclass[superscriptaddress,prd,onecolumn,nofootinbib]{revtex4-2}
\usepackage[utf8]{inputenc}

\usepackage[colorlinks=true, pdfstartview=FitV, linkcolor=red, citecolor=blue,
urlcolor=blue]{hyperref}
\usepackage[dvipdfmx]{graphicx}
\usepackage{amsmath,amssymb}
\usepackage{color}
\usepackage{slashed}
\usepackage[normalem]{ulem}

\newcommand{\diag}{{\rm{diag}}}

\newcommand{\sgn}{{\rm sgn}}

\newcommand{\ud}{\mathrm{d}}

\newcommand{\ue}{\mathrm{e}}

\renewcommand{\part}{{\rm part}}

\graphicspath{{./figures/}}

\begin{document}

\title{Chromomagnetic Condensate in Finite-Temperature SU(2) Yang-Mills Theory under Imaginary Rotation}

\author{Hao-Lei Chen}
\email{hlchen@shu.edu.cn}
\affiliation{Department of Physics, Shanghai University, Shanghai 200444, China}
\affiliation{Shanghai Research Center for Theoretical Nuclear Physics, NSFC and Fudan University, Shanghai 200438, China}
\author{Xu-Guang Huang}
\email{huangxuguang@fudan.edu.cn}
\affiliation{Physics Department and Center for Particle Physics and Field Theory, Fudan University, Shanghai 200438, China}
\affiliation{Key Laboratory of Nuclear Physics and Ion-beam Application (MOE), Fudan University, Shanghai 200433, China}
\affiliation{Shanghai Research Center for Theoretical Nuclear Physics, NSFC and Fudan University, Shanghai 200438, China}

\begin{abstract}

We investigate the finite-temperature SU(2) Savvidy model under an imaginary angular velocity. Employing the background-field method, we derive the one-loop effective potential and analyze both its real and imaginary parts. We demonstrate that imaginary rotation modifies the chromomagnetic condensate and the Polyakov loop, and can partially suppress the Nielsen–Olesen instability of the chromomagnetic background. Moreover, a high-temperature expansion shows that imaginary rotation strengthens the effective coupling and that the chromomagnetic field induces a negative contribution to the moment of inertia.
\end{abstract}

\maketitle


\section{Introduction}
Relativistic heavy-ion collisions provide a unique environment for exploring QCD matter under extreme conditions. In such collisions, the created fireball can carry a large vorticity~\cite{Betz:2007kg,Jiang:2016woz,Deng:2016gyh,Deng:2020ygd}. Experimental observations of spin polarization~\cite{STAR:2017ckg} have firmly established the presence of strong vortical fields in heavy-ion collisions and have triggered extensive theoretical efforts to understand the role of rotation and vorticity in QCD dynamics.

As a consequence, vorticity-related phenomena have attracted considerable attention, particularly in connection with hyperon and vector meson spin polarizations~\cite{Liang:2019clf,Gao:2020lxh,Huang:2020dtn,Liu:2020ymh,Becattini:2022zvf,Becattini:2024uha,Niida:2024ntm,Chen:2024afy}. Beyond spin physics, rotation has also been argued to influence the phase structure of strongly interacting matter. According to various model studies, rotation tends to restore chiral symmetry and favor deconfinement~\cite{Chen:2015hfc,Jiang:2016wvv,Chernodub:2016kxh,Chen:2021aiq,Fujimoto:2021xix,Chen:2023cjt,Zhao:2022uxc,Mameda:2023sst,Chen:2024jet,Zhu:2025pxh}. More recently, growing attention has been paid to the interplay between spin degrees of freedom and phase transitions in rotating QCD matter~\cite{Sun:2024anu,Xu:2022hql,Wei:2023pdf,Chen:2024hki}.

At the same time, lattice QCD studies of rotating systems have revealed several intriguing and sometimes puzzling features~\cite{Braguta:2021jgn,Yang:2023vsw}. In particular, the critical temperature of confinement/deconfinement phase transition has been found to increase with rotation, accompanied by a negative moment of inertia~\cite{Braguta:2023tqz}. These observations appear to be in qualitative tension with earlier model studies, which typically predict that rotation favors deconfinement and reduces the critical temperature.
Motivated by this apparent discrepancy, several attempts have been made to investigate Yang–Mills fields under rotation and to clarify the role of gluonic degrees of freedom~\cite{Chen:2022smf,Chen:2024tkr,Jiang:2023zzu,Jiang:2024zsw,Wang:2025mmv,Fukushima:2025hmh}. Nevertheless, the origin of the disagreement between theoretical approaches and lattice simulations, remains an open question.

One major obstacle in this context is the sign problem associated with real rotation in lattice simulations. From a theoretical perspective, real rotation also leads to a non–positive-definite elliptic operator in the Euclidean path integral, rendering the formulation of a well-defined thermal field theory highly nontrivial~\cite{Fursaev:2001yu}. A commonly adopted strategy to circumvent these difficulties is to study imaginary rotation, followed by an analytic continuation to real angular velocity at a later stage. This approach is closely analogous to the use of an imaginary chemical potential in finite-density QCD.
As a result, imaginary rotation provides a well-defined framework that enables direct comparison between continuum calculations and lattice simulations. Moreover, recent studies have shown that imaginary rotation itself is far from trivial: it can qualitatively modify the phase structure of pure Yang-Mills theories and even induce confinement-like behavior at high temperature~\cite{Chen:2022smf,Chen:2024tkr}.

Motivated by these developments, it is natural to revisit classic configurations of Yang–Mills theory in the presence of imaginary rotation. One such configuration is the Savvidy vacuum~\cite{Savvidy:1977as}, characterized by a constant chromomagnetic condensate. A well-known drawback of this configuration is the Nielsen–Olesen instability, which originates from tachyonic gluon modes~\cite{Nielsen:1978rm}. At finite temperature, the inclusion of a Polyakov-loop background modifies the thermal spectrum and the structure of the effective potential~\cite{Starinets:1994vi,Ebert:1996tj}. However, previous studies have shown that the Nielsen–Olesen instability generally persists even in the presence of a nontrivial Polyakov loop~\cite{Meisinger:2002ji,Bordag:2021vvw}, indicating that additional mechanisms are required to stabilize the system.
Nevertheless, the Savvidy model remains a valuable theoretical laboratory. In particular, the chromomagnetic condensate is directly related to the perturbative $\beta$ function of QCD and thus encodes essential information about asymptotic freedom and infrared dynamics~\cite{Nielsen:1980sx}. Moreover, since gluons are spin-1 particles, they are naturally susceptible to polarization effects, making chromomagnetic correlations especially relevant in rotating systems. From this perspective, studying how rotation—especially within the well-defined framework of imaginary rotation—modifies the Savvidy vacuum can provide useful insights into the interplay between confinement-related phenomena and rotational effects.

In this work, we therefore study the finite-temperature SU(2) Savvidy model in the presence of an imaginary angular velocity, incorporating simultaneously a constant chromomagnetic condensate and a Polyakov-loop background. By working directly with imaginary rotation, we circumvent the technical complications arising from the combined presence of magnetic fields and real rotation, and provide a well-defined framework suitable for comparison with lattice studies. Our goal is to clarify how imaginary rotation modifies the effective potential, the chromomagnetic condensate, and the Polyakov-loop background. In particular, we extract the dependence of the effective coupling constant on the imaginary angular velocity, as well as the nontrivial dependence of the moment of inertia on the chromomagnetic condensate.

The paper is organized as follows.
Section~\ref{sec:setup} introduces the setup and background fields.
The one-loop effective potential is derived in Sec.~\ref{sec:V}, with numerical results presented in Sec.~\ref{sec:numerical}.
A small–imaginary-angular-velocity expansion is performed in Sec.~\ref{sec:expansion}, from which the effective coupling constant and the moment of inertia are extracted and their physical implications are discussed.
Conclusions and discussions are given in Sec.~\ref{sec:discussion}.

\section{setup}\label{sec:setup}
We start with the Euclidean metric tensor describing a system under an imaginary angular velocity $\Omega_I$,
\begin{equation}
g_{\mu\nu}
=
\begin{pmatrix}
-1-\Omega_I^{2} r^{2} & y\,\Omega_I & -x\,\Omega_I & 0 \\
y\,\Omega_I & -1 & 0 & 0 \\
- x\,\Omega_I & 0 & -1 & 0 \\
0 & 0 & 0 & -1
\end{pmatrix}.
\end{equation}
To obtain the effective potential in the presence of a chromomagnetic condensate and a Polyakov-loop background under imaginary rotation, we employ the background-field method. The gauge field is decomposed as $\mathcal{A}^a_\mu=\bar A^a_\mu+A^a_\mu$, where $\bar A^a_\mu$ denotes the background field and $A^a_\mu$ represents quantum fluctuations. The Euclidean Lagrangian then reads
\begin{equation}
\mathcal{L}_E
=
\frac{1}{4}\,\bar F^a_{\mu\nu}\bar F^{a\mu\nu}
+\frac{1}{2}\,\bar F^a_{\mu\nu} F^{a\mu\nu}
+\frac{1}{4}\,F^a_{\mu\nu} F^{a\mu\nu}
+\mathcal{L}_{\mathrm{gf}}
+\mathcal{L}_{\mathrm{gh}} ,
\end{equation}
with the gauge-fixing term (we adopt the Feynman gauge throughout this work)
\begin{equation}
\mathcal{L}_{\mathrm{gf}}
= \frac{1}{2}\bigl(D_\mu^{B} A^{a\mu}\bigr)^2 ,
\end{equation}
and the ghost term
\begin{equation}
\mathcal{L}_{\mathrm{gh}}
=
\bar c^{\,a}\, D_\mu^{B\,ab}\, D^{\mu\,bc}\, c^{\,c}.
\end{equation}
The field-strength tensors and background-covariant derivatives are defined as
\begin{align}
\bar F^a_{\mu\nu}
&= \nabla_\mu \bar A^a_\nu - \nabla_\nu \bar A^a_\mu
   - g f^{abc}\, \bar A^b_\mu \bar A^c_\nu , \\[4pt]
F^a_{\mu\nu}
&= D_\mu^{B} A^a_\nu - D_\nu^{B} A^a_\mu
   - g f^{abc}\, A^b_\mu A^c_\nu , \\[4pt]
D_\mu^{B} A^a_\nu
&= \bigl(\nabla_\mu \delta^{ac} - g f^{abc}\, \bar A^b_\mu \bigr) A^c_\nu , \\[4pt]
D_\mu c^a
&= \bigl[\nabla_\mu \delta^{ac}
   - g f^{abc}\,(\bar A_\mu^b + A_\mu^b)\bigr] c^c ,
\end{align}
where $\nabla_\mu$ denotes the spacetime covariant derivative, $f^{abc}$ are the structure constants, and repeated color indices are summed over.

It is convenient to work in the tangent space, which is related to the coordinate space by the vierbein
\begin{equation}
e_{\mu}^{\ \hat{\mu}}
=
\begin{pmatrix}
1 & 0 & 0 & 0 \\
- y\,\Omega_I & 1 & 0 & 0 \\
\;\;x\,\Omega_I & 0 & 1 & 0 \\
0 & 0 & 0 & 1
\end{pmatrix}.
\end{equation}
Here we use indices with hat to denote tangent space. The vierbein is chosen to satisfy the condition $g_{\mu\nu}=e_{\phantom{1}\mu}^{\hat\mu}e_{\phantom{1}\nu}^{\hat\nu}\eta_{\hat\mu\hat\nu}$, where $\eta_{\hat\mu\hat\nu}=\diag(-1,-1,-1,-1)$ is the non-rotating metric. The covariant derivative can then be written as
\begin{equation}
    \nabla _\mu A^a_\nu=e_{\phantom{1}\nu}^{\hat\nu}(\partial_\mu A^a_{\hat \nu}-\omega_{\mu\phantom{1}\hat \nu}^{\phantom{1}\hat \lambda}A^a_{\hat\lambda}),
\end{equation}
where $\omega_{\mu\phantom{1}\hat \nu}^{\phantom{1}\hat \lambda}$ is the spin connection with nonzero components $\omega_{t\phantom{1}\hat x}^{\phantom{1}\hat y}=-\omega_{t\phantom{1}\hat y}^{\phantom{1}\hat x}=\Omega_I$.

In this work, we consider both a Polyakov-loop condensate $\phi$ and a chromomagnetic field $H$. We assume that the background gauge field in the tangent space takes the form
$\bar A^a_{\hat\mu}=\delta^a_3 \bar A^3_{\hat\mu}$, with
\begin{equation}
		\bar A^3_{\hat\mu}=(\phi,\frac{1}{2}Hy,-\frac{1}{2}Hx,0),
\end{equation}
so that only the Abelian component is nonvanishing and the chromomagnetic field is aligned with the rotation axis. 
Since rotation already selects a preferred direction, this choice is physically natural and simplifies the analysis. We further assume that spatial derivatives of the chromomagnetic field $gH$ and the Polyakov-loop background $\phi$ can be neglected (local density approximation), since our analysis focuses on the region near the system center ($r=0$), where spatial gradients of the background fields give only subleading contributions.

Using the vierbein formalism, the Lagrangian can be expressed explicitly in the tangent space. After straightforward calculations, we obtain
\begin{equation}
\begin{aligned}
\mathcal{L}_E
=&\;
\frac{1}{2} H^2
+ A_-^{+}\,
\Big[
- \big( \partial_\tau - i \Omega_I \hat L_z + i \Omega_I + i g \phi \big)^2
- \partial_i^2
- g H \hat L_z
+ \frac{1}{4} g^2 H^2 r^2
+ 2 g H
\Big]\,
A_+^{-}
\\[6pt]
&\;
+ A_+^{+}\,
\Big[
- \big( \partial_\tau - i \Omega_I \hat L_z - i \Omega_I + i g \phi \big)^2
- \partial_i^2
- g H \hat L_z
+ \frac{1}{4} g^2 H^2 r^2
- 2 g H
\Big]\,
A_-^{-},
\end{aligned}
\end{equation}
where we define
\begin{equation}
\begin{aligned}
A_{\pm}^a
&= \frac{1}{\sqrt{2}}
\left( A_{\hat x}^a \pm A_{\hat y}^a \right), \\[4pt]
A_{\pm}^{+}
&= \frac{1}{\sqrt{2}}
\left( A_{\pm}^1 + A_{\pm}^2 \right),
\\[4pt]
A_{\pm}^{-}
&= \frac{1}{\sqrt{2}}
\left( A_{\pm}^1 - A_{\pm}^2 \right).
\end{aligned}
\end{equation}
We omit the contribution from the neutral field $A^3_{\hat\mu}$ in Secs.~\ref{sec:V} and \ref{sec:numerical}, since it does not couple to the background fields. Nevertheless, it remains sensitive to rotation and contributes to the total angular momentum of the system; its contribution will therefore be recovered when we discuss the moment of inertia in Sec.~\ref{sec:expansion}. Unphysical modes are already canceled by the ghost fields.

\section{Evaluation of the effective potential}\label{sec:V}
Within the local density approximation, the one-loop effective potential can be evaluated using standard imaginary-time thermal field theory,
\begin{equation}
\begin{aligned}
V(r)
=&\;
\frac{1}{2} H^2
+ \sum_{n=-\infty}^{\infty}
  \sum_{\lambda=0}^{\infty}
  \sum_{l=-\lambda}^{\,N-\lambda}
  \sum_{s=\pm 1}
  \int \frac{\mathrm{d}k_z}{2\pi}
\\
&\times
\ln\!\Big[
\big( \omega_n - \Omega_I \big( \sgn(gH)\, l - s \big) + g \phi \big)^2
+ |gH| \big( 2\lambda + 1 + 2s \big)
+ k_z^2
\Big]
\\
&\times
\frac{|gH|}{2\pi}\,
\Phi_l^{\,2}
\!\left(
\lambda,\,
\frac{1}{2} |gH| r^2
\right).
\end{aligned}
\end{equation}
Here $\omega_n=2\pi n/\beta$ with $\beta=1/T$ are the bosonic Matsubara frequencies, and
$N=\lfloor |gH|S/2\pi\rfloor$ denotes the Landau-level degeneracy with $S$ the transverse area of the system. The corresponding eigenfunctions are
\begin{equation}
\Phi_{l}(\lambda,x)
=
\left[
\frac{\lambda!}{(\lambda+l)!}
\right]^{\frac{\mathrm{sgn}(l)}{2}}
x^{\frac{|l|}{2}}
\mathrm{e}^{-x/2}\,
L^{|l|}_{\,\lambda-(|l|-l)/2}(x),
\end{equation}
where $L^a_b(x)$ is the associated Laguerre polynomial. The eigenfunctions satisfy the normalization condition
\begin{equation}
\frac{|gH|}{2\pi}
\int_{0}^{\infty} r\,\mathrm{d}r\;
\Phi_{l}^{\,2}
\!\left(
\lambda,\,
\frac{1}{2}|gH|\,r^{2}
\right)
= 1 .
\end{equation}
Note that the direction of the chromomagnetic field enters the effective potential through its coupling to the orbital angular-momentum quantum number $l$.

In this work, we focus on the system center $r=0$, following Ref.~\cite{Chen:2022smf}. Away from the center, orbital angular momentum contributions become important, and analytical calculations are considerably more involved because the summation is intertwined with the Landau-level index $\lambda$. Even in the absence of a chromomagnetic background, the Polyakov-loop condensate exhibits a nontrivial radial dependence~\cite{Chen:2024tkr}. We therefore defer a detailed analysis of the full radial dependence to future work.

At the system center, only modes with $l=0$ contribute. As a result, the effective potential is independent of the direction of the chromomagnetic field, and we may restrict to $gH>0$ without loss of generality.
 In this case, the potential reduces to
\begin{equation}\label{eq:V0nosum}
\begin{aligned}
V(r=0)
=&\;
\frac{H^2}{2} 
+ \frac{gH}{2\pi\beta}
\sum_{n=-\infty}^{\infty}
\sum_{\lambda=0}^{\infty}
\sum_{s=\pm 1}
\int \frac{\mathrm{d}k_z}{2\pi}
\\
&\times
\ln\!\Big[
\big( \omega_n + s\,\Omega_I + g\phi \big)^2
+ gH \big( 2\lambda + 1 + 2s \big)
+ k_z^2
\Big].
\end{aligned}
\end{equation}
The imaginary angular velocity enters as a spin-dependent chemical potential, in agreement with Ref.~\cite{Chen:2022smf}.
It is evident from Eq.~(\ref{eq:V0nosum}) that modes with $\lambda=0$ and $s=-1$ can lead to a negative argument of the logarithm for $n=0$ and sufficiently small $k_z$, resulting in a nonvanishing imaginary part of the effective potential. This signals an instability of the constant chromomagnetic configuration, first identified by Nielsen and Olesen~\cite{Nielsen:1978rm}. In the present setup, the chromomagnetic configuration is stable provided that the effective potential has no imaginary part, which is guaranteed if the condition
\begin{equation}\label{eq:stablilty_condition}
    (\omega_n+g\phi-\Omega_I)^2\geq gH
\end{equation}
is satisfied for all Matsubara frequencies. The case without rotation has been analyzed in Ref.~\cite{Meisinger:2002ji}, where it was shown that the minimum of the effective potential is always accompanied by a nonzero imaginary part. In particular, at high temperature the minimum occurs at $\phi=0$ in the absence of rotation. By contrast, imaginary rotation can induce a nonzero Polyakov-loop condensate $\phi\neq0$~\cite{Chen:2022smf}, making it natural to ask whether imaginary rotation can modify—or possibly suppress—the instability of the chromomagnetic background.

In the following, we refer to the mode with $\lambda=0$ and $s=-1$ as the tachyonic mode, while the remaining modes are referred to as non-tachyonic. We begin by evaluating the contribution of the non-tachyonic modes to the effective potential, denoted by $V_{nt}$,
\begin{equation}
V_{nt}
=
\frac{gH}{2\pi\beta}
\sum_{n=-\infty}^{\infty}
\sum_{\mathrm{n.t.}}
\int \frac{\mathrm{d}k_z}{2\pi}
\ln\!\Big[
\big( \omega_n + s\,\Omega_I + g\phi \big)^2
+ gH \big( 2\lambda + 1 + 2s \big)
+ k_z^2
\Big].
\end{equation}
where $nt$ stands for non-tachyonic modes. The logarithm can be represented using the Schwinger proper-time formalism,
\begin{eqnarray}
    \ln A=-\int_0^\infty\frac{\ud t}{t}\ue^{-tA}.
\end{eqnarray}
Using this representation, $V_{nt}$ can be rewritten as
\begin{equation}
\begin{aligned}
V_{nt}
=&\;
-\frac{gH}{2\pi\beta}
\sum_{n=-\infty}^{\infty}
\sum_{\mathrm{n.t.}}
\int \frac{\mathrm{d}k_z}{2\pi}
\int_{0}^{\infty} \frac{\mathrm{d}t}{t}\,
\exp\!\Big\{
- t \big[
(\omega_n + s\,\Omega_I + g\phi)^2
+ gH ( 2\lambda + 1 + 2s )
+ k_z^2
\big]
\Big\}
\\[6pt]
=&\;
-\frac{gH}{8\pi^2}
\sum_{\mathrm{n.t.}}
\int_{0}^{\infty} \frac{\mathrm{d}t}{t^{2}}\,
\exp\!\big[ - t\, gH ( 2\lambda + 1 + 2s ) \big]
\sum_{n=-\infty}^{\infty}
\exp\!\left( - \frac{n^{2}\beta^{2}}{4t} \right)
\cos\!\big[ n\beta ( g\phi + s\,\Omega_I ) \big].
\end{aligned}
\end{equation}
In deriving the second line, we have performed the integration over $k_z$ and used the identity
\begin{equation}
\sum_{n=-\infty}^{\infty}
\exp\!\big[ - t ( 2\pi n T + x )^{2} \big]
=
\frac{\beta}{2\sqrt{\pi t}}
\sum_{n=-\infty}^{\infty}
\exp\!\left( - \frac{n^{2}\beta^{2}}{4t} \right)
\cos\!\big( n\beta x \big).
\end{equation}
After performing the summation over $\lambda$ and $s$ for all non-tachyonic modes, we obtain
\begin{equation}\label{eq:Vnt_sum}
\begin{aligned}
V_{nt}
=&\;
-\frac{gH}{8\pi^{2}}
\int_{0}^{\infty} \frac{\mathrm{d}t}{t^{2}}
\sum_{n=-\infty}^{\infty}
\exp\!\left( - \frac{n^{2}\beta^{2}}{4t} \right)
\frac{1}{\mathrm{e}^{2t gH}-1}
\\
&\times
\Big[
\mathrm{e}^{-t gH}
\cos\!\big( n\beta ( g\phi + \Omega_I ) \big)
+
\mathrm{e}^{\,t gH}
\cos\!\big( n\beta ( g\phi - \Omega_I ) \big)
\Big].
\end{aligned}
\end{equation}
In the zero-temperature limit, only the $n=0$ term survives, corresponding to the vacuum contribution, which is independent of $\phi$ and $\Omega_I$.
 We denote this vacuum term as $V^{T=0}_{nt}$. Our interest is in the finite-temperature part, $V^T_{nt}$.
Using the Jacobi theta function,
\begin{equation}
    \vartheta_3(z,q) = 1 + 2 \sum_{n=1}^{\infty} q^{n^2}\cos(2 n z),
\end{equation}
we arrive at the following representation:
\begin{equation}\label{eq:Vnt}
\begin{aligned}
V_{nt}^{T}
=&\;
-\frac{gH}{8\pi^{2}}
\int_{0}^{\infty} \frac{\mathrm{d}t}{t^{2}}\,
\frac{1}{\mathrm{e}^{2t gH}-1}
\Bigg[
\mathrm{e}^{-t gH}
\left(
\vartheta_{3}
\!\left(
\frac{\beta ( g\phi + \Omega_{I} )}{2},
\,\mathrm{e}^{-\frac{\beta^{2}}{4t}}
\right)
- 1
\right)
\\
&\qquad
+ \mathrm{e}^{t gH}
\left(
\vartheta_{3}
\!\left(
\frac{\beta ( g\phi - \Omega_{I} )}{2},
\,\mathrm{e}^{-\frac{\beta^{2}}{4t}}
\right)
- 1
\right)
\Bigg].
\end{aligned}
\end{equation}

Having obtained the contribution from the non-tachyonic modes, we now turn to the tachyonic sector,
\begin{equation}
\begin{aligned}
V_{ta}
=&\;
\frac{gH}{2\pi\beta}
\sum_{n=-\infty}^{\infty}
\int \frac{\mathrm{d}k_z}{2\pi}\,
\ln\!\Big[
\big( \omega_n - \Omega_I + g\phi \big)^2
- gH
+ k_z^2
\Big]
\\[6pt]
=&\;
\frac{gH}{2\pi\beta}
\int \frac{\mathrm{d}k_z}{2\pi}
\Big[
\beta\,\varepsilon_{ta}
+ \ln\!\big( 1 - \mathrm{e}^{-\beta(\varepsilon_{ta}- i g\phi + i\Omega_I)} \big)
+ \ln\!\big( 1 - \mathrm{e}^{-\beta(\varepsilon_{ta}+ i g\phi - i\Omega_I)} \big)
\Big],
\end{aligned}
\end{equation}
where we define $\varepsilon_{ta}\equiv \sqrt{k_z^{2}-gH}$.
The first term in brackets is the zero-temperature contribution, which we denote by $V^{T=0}_{{ta}}$. The remaining terms define the finite-temperature part $V^{T}_{ta}$, which requires careful treatment.
To isolate the branch cut of the logarithm and to facilitate numerical calculations, we separate the $k_z$-integration into two parts,
$|k_z|<\sqrt{gH}$ and $|k_z|>\sqrt{gH}$.
The former yields an imaginary contribution and must be evaluated on the principal branch, whereas the latter is purely real.
Accordingly, we write
$
V^{T}_{{ta}} = V^{T,<}_{{ta}} + V^{T,>}_{{ta}},
$
with $V^{T,<}_{{ta}}$ and $V^{T,>}_{{ta}}$ denoting the contributions from $|k_z|<\sqrt{gH}$ and $|k_z|>\sqrt{gH}$, respectively.

For $|k_z|>\sqrt{gH}$, the contribution can be written as
\begin{equation}\label{eq:Vtag}
V_{ta}^{T,>}
=
\frac{gH}{\pi\beta}
\int_{\sqrt{gH}}^{\infty}
\frac{\mathrm{d}k_z}{2\pi}\,
\ln\!\Big[
1
- 2\,\mathrm{e}^{-\beta \varepsilon_{ta}}
\cos\!\big( \beta ( g\phi - \Omega_I ) \big)
+ \mathrm{e}^{-2\beta \varepsilon_{ta}}
\Big],
\end{equation}
which is purely real and numerically well behaved. While for the region $|k_z|<\sqrt{gH}$,
we apply the $i\epsilon$ prescription $gH\to gH+i\epsilon$ for the analytic continuation
\begin{equation}
    \sqrt{k_z^2-gH-i\epsilon}=-i\tilde\varepsilon_{ta},
\end{equation}
where we define $\tilde\varepsilon_{ta}=\sqrt{gH-k_z^{2}}$.
Then the real part reads
\begin{equation}\label{eq:ReVtal}
\Re\, V_{ta}^{T,<}
=
\frac{gH}{\pi\beta}
\int_{0}^{\sqrt{gH}}
\frac{\mathrm{d}k_z}{2\pi}\,
\Big\{
\ln\!\big[
2 - 2 \cos\!\big( \beta ( \tilde\varepsilon_{ta} - g\phi + \Omega_I ) \big)
\big]
+
\ln\!\big[
2 - 2 \cos\!\big( \beta ( \tilde\varepsilon_{ta} + g\phi - \Omega_I ) \big)
\big]
\Big\}.
\end{equation}
 To evaluate the imaginary part explicitly, we use the principal-branch identity 
\begin{equation}
    \Im\ln(1-e^{ix})=\frac{1}{2}(\mathrm{mod}(x,2\pi)-\pi),
\end{equation}
which defines the phase continuously in the interval $x\bmod2\pi\in(0,2\pi)$. The imaginary part then becomes
\begin{equation}\label{eq:ImVtal}
\begin{split}
\Im\,V_{ta}^{T,<}
&=\frac{gH}{\pi\beta}\int_{0}^{\sqrt{gH}}\!\frac{\ud k_z}{2\pi}\,
\Im\!\Big[
\ln\!\big(1-e^{i\beta(\tilde\varepsilon_{ta}-g\phi+\Omega_I)}\big)+\ln\!\big(1-e^{i\beta(\tilde\varepsilon_{ta}+g\phi-\Omega_I)}\big)
\Big]\\[4pt]
&=\,\frac{gH}{2\pi\beta}\int_{0}^{\sqrt{gH}}\!\frac{\ud k_z}{2\pi}\,
\Big\{
\mathrm{mod}\!\big[\beta(\tilde\varepsilon_{ta}-g\phi+\Omega_I),2\pi\big]
+\mathrm{mod}\!\big[\beta(\tilde\varepsilon_{ta}+g\phi-\Omega_I),2\pi\big]
-2\pi
\Big\}.
\end{split}
\end{equation}
Finally, combining Eqs.~\eqref{eq:Vnt},~\eqref{eq:Vtag},~\eqref{eq:ReVtal},~\eqref{eq:ImVtal}, together with the vacuum contribution extensively discussed in the literature~\cite{Nielsen:1978rm,Meisinger:2002ji,greiner2002quantum},
\begin{equation}
    V^{T=0}=\frac{H^2}{2}+\frac{11(gH)^2}{48\pi^2}\ln\frac{gH}{\mu^2}-i\frac{(gH)^2}{8\pi},
\end{equation}
where $\mu$ is a renormalization-group invariant scale, we obtain the full effective potential at finite temperature
\begin{equation}\label{eq:Veff_total}
\begin{split}
    V_R\equiv&\Re V=\Re V^{T=0}+V_{nt}^T+V_{ta}^{T,>}+\Re V_{ta}^{T,<},\\
    V_I\equiv&\Im V=\Im V^{T=0}+\Im V_{ta}^{T,<}.
\end{split}
\end{equation}

 \begin{figure}[tp]
\begin{minipage}[t]{0.45\linewidth}
\includegraphics[width=1\textwidth]{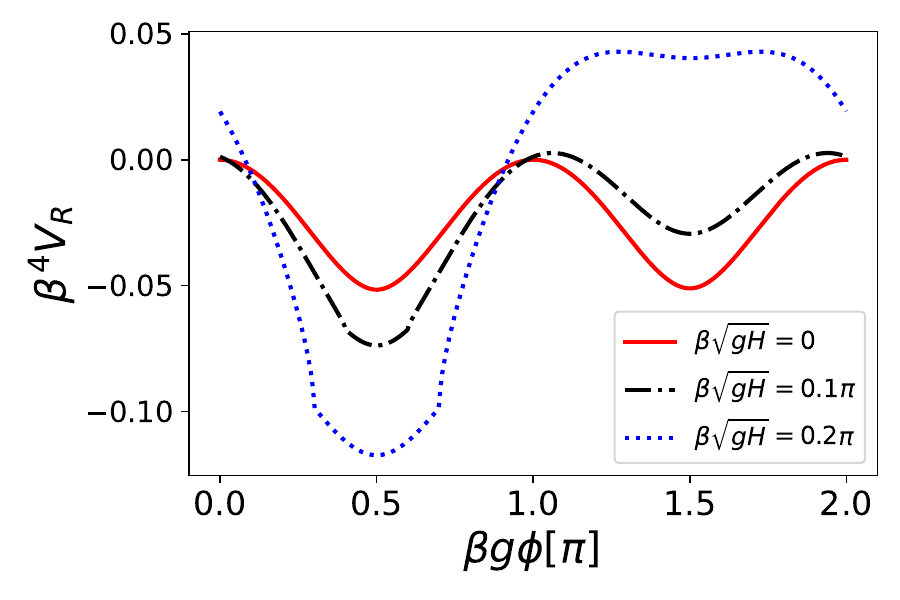}
\caption{Real part of the effective potential $V_R$ as a function of $\beta g\phi$ for several values of $\beta\sqrt{gH}$ at $\Omega_I=\pi/2$.}
\label{fig:Veff_phi}
\end{minipage}%
\hfill
\end{figure}

To gain further insight into the structure of the effective potential, we analyze two limiting cases. Specifically, we consider
(i) a finite chromomagnetic field ($gH\neq0$) with vanishing Polyakov-loop background and imaginary rotation ($\phi=\Omega_I=0$),
and (ii) the limit in which the chromomagnetic condensate disappears, $gH\to0$.

We first consider the case of a finite chromomagnetic field with $\phi=\Omega_I=0$. In the high-temperature regime, $\beta\sqrt{gH}\ll2\pi$, the imaginary part of the effective potential becomes
\begin{equation}
\begin{split}
    V_I
    &= -\,\frac{(gH)^2}{8\pi}
    + \frac{gH}{2\pi\beta}
    \int_{0}^{\sqrt{gH}}\!\frac{dk_z}{2\pi}\,
    (2\beta\sqrt{gH-k_z^2}-2\pi) \\[4pt]
    &= -\,\frac{(gH)^{3/2}}{2\pi\beta},
\end{split}
\end{equation}
in agreement with Refs.~\cite{Ninomiya:1981eq,Persson:1996zy}.

We next turn to the limit $gH\to0$, in which the chromomagnetic condensate vanishes. In this case, only the non-tachyonic contribution survives, yielding
\begin{equation}
\begin{split}
    V_{nt}^{T}
    &= -\,\frac{1}{4\pi^2} \sum_{n=1}^{\infty}\int_{0}^{\infty}\!\frac{dt}{2t^3}\,\exp\!\left(-\frac{n^2\beta^2}{4t}\right)\big[\cos n\beta(g\phi+\Omega_I)+\cos n\beta(g\phi-\Omega_I)\big] \\[4pt]
    &= -\,\frac{2}{\pi^2\beta^4}
    \sum_{n=1}^{\infty}
    \frac{\cos n\beta(g\phi+\Omega_I)+\cos n\beta(g\phi-\Omega_I)}{n^4} \\[4pt]
    &= -\,\frac{1}{\pi^2\beta^4}\!
    \left[
    \mathrm{Li}_4(e^{i\beta(g\phi+\Omega_I)})
    + \mathrm{Li}_4(e^{-i\beta(g\phi+\Omega_I)})
    + \mathrm{Li}_4(e^{i\beta(g\phi-\Omega_I)})
    + \mathrm{Li}_4(e^{-i\beta(g\phi-\Omega_I)})
    \right],
\end{split}
\end{equation}
which exactly reproduces Eq.~(9) of Ref.~\cite{Chen:2022smf}.
If we further set $\Omega_I=0$ and restrict to $g\phi\in[0,2\pi)$, this expression simplifies to
\begin{equation}
\begin{split}
    V_{nt}^{T}
    &= -\,\frac{4}{\pi^2\beta^4}
    \left[
    \frac{\pi^4}{90}
    - \frac{\pi^2(\beta g\phi)^2}{12}
    + \frac{\pi(\beta g\phi)^3}{12}
    - \frac{(\beta g\phi)^4}{48}
    \right],
\end{split}
\end{equation}
which is the well-known Gross–Pisarski–Yaffe (GPY) or Weiss potential~\cite{Weiss:1980rj,Weiss:1981ev,Gross:1980br,Fukushima:2017csk}.

\section{numerical results}\label{sec:numerical}
Using the full effective potential in Eq.~\eqref{eq:Veff_total}, we now perform numerical calculations to explore its behavior at high temperature in the presence of imaginary rotation.
Fig.~\ref{fig:Veff_phi} shows the real part of the effective potential, $V_R$, as a function of the Polyakov-loop phase $\beta g\phi$ for several values of $\beta\sqrt{gH}$ at $\Omega_I=\pi/2$.
In the absence of a chromomagnetic field, the effective potential exhibits two degenerate minima, reflecting the underlying $\mathbb{Z}_2$ center symmetry.
Once a finite chromomagnetic field is introduced, this degeneracy is lifted: one minimum is energetically favored, while the other is pushed upward.
This asymmetry originates from the contribution of the lowest Landau level ($\lambda=0$, $s=-1$) and the next-to-lowest Landau level ($\lambda=1$, $s=-1$), which explicitly break the $\mathbb{Z}_2$ symmetry of the effective potential.

For the largest chromomagnetic field shown in Fig.~\ref{fig:Veff_phi} (blue dotted curve), the real part of the effective potential develops two cusps near the minimum.
These cusps correspond to points where the first derivative of $V_R$ is discontinuous. The interval between the two non-differentiable points coincides with the region in which the effective potential acquires an imaginary part, reflecting the Nielsen--Olesen instability and the emergence of tachyonic modes associated with the lowest Landau level.

We now determine the equilibrium configuration by minimizing the real part of the effective potential with respect to the dynamical variables $g\phi$ and $gH$,
\begin{equation}
    \frac{\partial V_R}{\partial (gH)}=\frac{\partial V_R}{\partial (g\phi)}=0.
\end{equation}
At this stage we restrict our analysis to the real part of the effective potential.
The impact of imaginary rotation on the imaginary component of the potential will be addressed later in this section.

\begin{figure}[t]
\begin{minipage}[t]{0.45\linewidth}
\includegraphics[width=1\columnwidth]{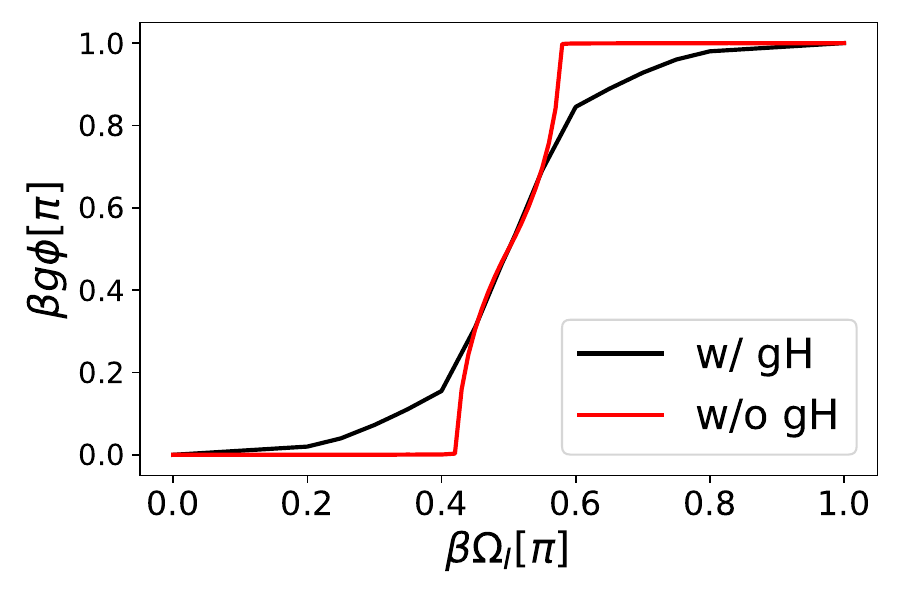}
\caption{Polyakov-loop phase $\beta g\phi$ as a function of the imaginary angular velocity $\Omega_I$ at high temperature $T=10\mu$.  
The black curve corresponds to the case with a chromomagnetic condensate $gH$, while the red curve shows the result without taking into account the condensate.
}\label{fig:phi_omega}
\end{minipage}%
\hfill
\begin{minipage}[t]{0.45\linewidth}
\includegraphics[width=1\columnwidth]{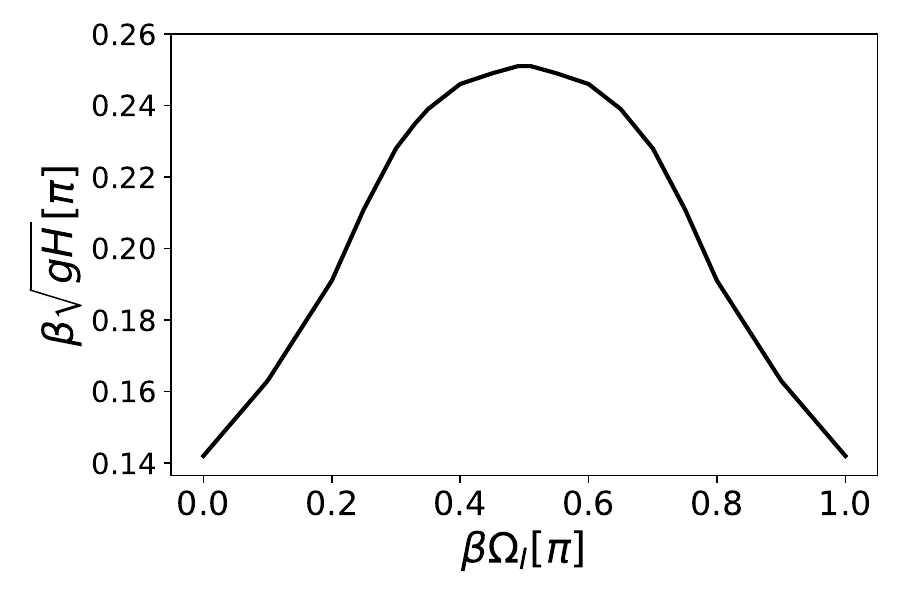}
    \caption{Dependence of the chromomagnetic condensate on the imaginary angular velocity $\Omega_I$ at fixed temperature $T = 10\mu$.}
    \label{fig:gH_omega}
\end{minipage}
\end{figure}

Fig.~\ref{fig:phi_omega} and Fig.~\ref{fig:gH_omega} show the dependence of the Polyakov-loop phase and the chromomagnetic condensate on the imaginary angular velocity $\Omega_I$, respectively, at high temperature ($T=10\mu$).
For comparison, Fig.~\ref{fig:phi_omega} also includes the result obtained in the absence of the chromomagnetic condensate~\cite{Chen:2022smf} (red curve).
At $\Omega_I=0$, the Polyakov-loop condensate vanishes, indicating a deconfined phase in which the $\mathbb{Z}_2$ center symmetry is spontaneously broken.
Once a finite imaginary rotation is introduced, however, the $\mathbb{Z}_2$ symmetry is explicitly broken by the combined effect of $\Omega_I$ and the chromomagnetic condensate.
As a result, $\beta g\phi$ immediately deviates from zero even in the small-$\Omega_I$ region.
A similar behavior appears near $\Omega_I\lesssim\pi$, where the preferred value of $\beta g\phi$ shifts slightly away from the center-symmetric point $\beta g\phi=\pi$.

On the other hand, the chromomagnetic condensate increases with imaginary rotation and reaches its maximum at $\beta\Omega_I=\pi/2$.
Notably, $gH(\Omega_I)$ is symmetric about this point.
This symmetry follows from the invariance of the effective potential—both its real and imaginary parts—under the transformation
\begin{equation}
    \beta g\phi \rightarrow \pi - \beta g\phi, \quad\beta\Omega_I \rightarrow \pi -\beta\Omega_I,
\end{equation}
which is evident from Eq.~\eqref{eq:V0nosum}.

\begin{figure}[t]
\begin{minipage}[t]{0.45\linewidth}
\includegraphics[width=1\columnwidth]{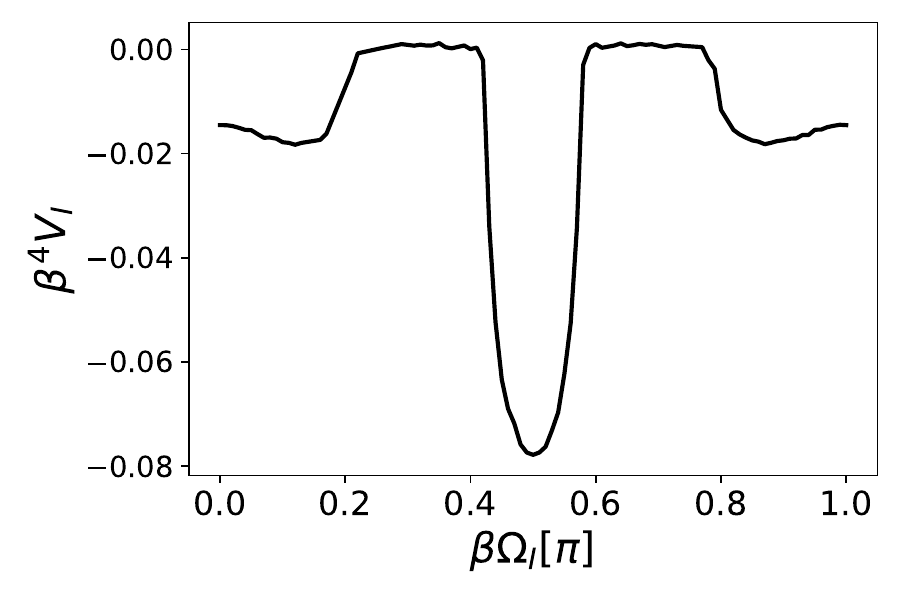}
\caption{Imaginary part of the effective potential, $V_I$, as a function of the imaginary angular velocity $\Omega_I$ at temperature $T=10\mu$, evaluated at the corresponding minima of the real part $V_R$.
}\label{fig:VI}
\end{minipage}%
\hfill
\begin{minipage}[t]{0.45\linewidth}
\includegraphics[width=1\columnwidth]{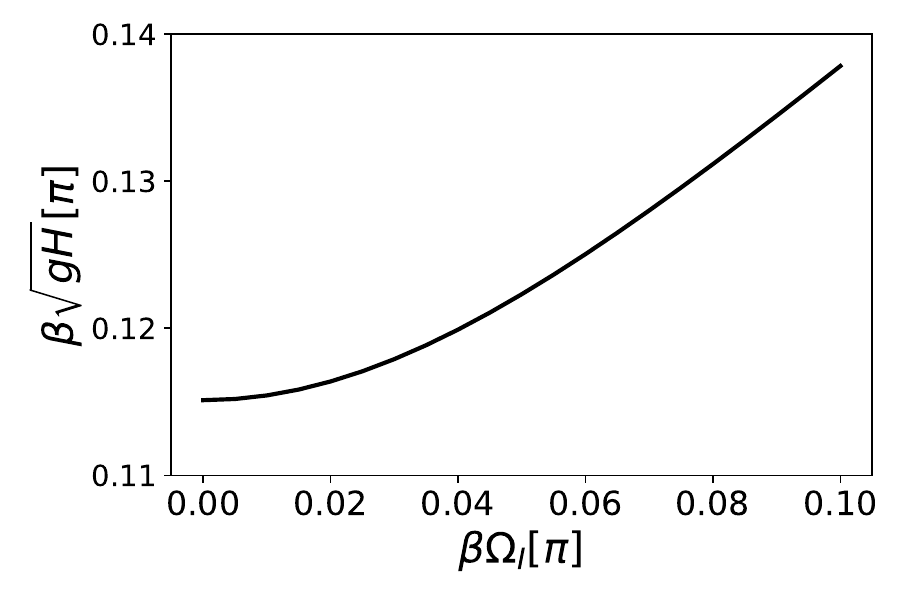}
\caption{
Dependence of the chromomagnetic condensate on the imaginary angular velocity $\Omega_I$
at fixed temperature $T = 10\mu$, obtained from the high-temperature expansion Eq.~\eqref{eq:VRexpansion}.
}
    \label{fig:small_omega}
\end{minipage}
\end{figure}

We next examine the behavior of the imaginary part of the effective potential evaluated at the corresponding minima of $V_R$.
As shown in Ref.~\cite{Meisinger:2002ji}, the inclusion of a nontrivial Polyakov-loop background in the Savvidy model does not remove the Nielsen--Olesen instability; an imaginary contribution to the effective potential generally persists.
However, in the present setup, the system is additionally subjected to an imaginary angular velocity, which acts as a spin-dependent chemical potential.
This extra control parameter opens up the possibility that the instability is suppressed.
Indeed, Fig.~\ref{fig:VI} shows the dependence of the imaginary part of the effective potential, $V_I$, on the imaginary angular velocity $\Omega_I$.
We find that there exists a finite interval,
approximately $0.2 \lesssim \beta\Omega_I \lesssim 0.4$,
within which $V_I$ vanishes, indicating a stable configuration.
This stable window coincides with the cusp structure observed in Fig.~\ref{fig:phi_omega}.

\section{Effective coupling constant and moment of inertia}
\label{sec:expansion}

Our numerical results show that in the small--$\Omega_I$ region the Polyakov-loop condensate remains very small.  
This observation allows us to set $g\phi \simeq 0$ in this regime without introducing a significant error.  
Under this assumption, the effective potential can be systematically expanded in powers of the imaginary angular velocity.  
In particular, retaining terms up to $O(\Omega_I^2)$ and $O(g^2H^2)$ allows us to extract the nontrivial dependence of the effective coupling constant on the imaginary angular velocity.

For $g\phi = 0$, it is advantageous to express the effective potential in an alternative form, which we shall employ for the high-temperature expansion below,
\begin{equation}\label{eq:VhighT}
\begin{aligned}
V(r=0)
=&\;
V_{\Omega_I}
+\frac{H^2}{2} + \frac{11 g^{2} H^{2}}{48\pi^{2}}
  \ln\!\left( \frac{gH}{\mu^{2}} \right)
- \frac{11}{24\pi^{2}} (gH)^{2}
  \left[
    \ln\!\left( \frac{\beta \sqrt{gH}}{4\pi} \right)
    + \gamma
  \right]
+ \frac{7}{32\pi^{2}} (gH)^{2}
+ C_{1}\, \frac{(gH)^{2}}{8\pi^{2}}
\\[4pt]
&\;
- \frac{1}{\pi \beta^{4}}
  {\sum_{l}}^\prime
  \Bigg\{
  \frac{1}{3}
  \Big[ \beta^{2} gH + ( \beta \Omega_I - 2\pi l )^{2} \Big]^{3/2}
  - \frac{1}{3} \big| \beta \Omega_I - 2\pi l \big|^{3}
  - \frac{\beta^{4} (gH)^{2}}{16\pi |l|}
  \Bigg\}
\\[4pt]
&\;
- \frac{(gH)^{2}}{12\pi}
  {\sum_{l}}^\prime
  \left[
  \frac{1}{\sqrt{ \beta^{2} gH + ( \beta \Omega_I - 2\pi l )^{2} }}
  - \frac{1}{2\pi |l|}
  \right]
\\[4pt]
&\;
+ \frac{gH}{2\pi \beta^{2}}
  {\sum_{l}}^\prime
  \left[
  \sqrt{ - \beta^{2} gH + ( \beta \Omega_I - 2\pi l )^{2} }
  + \frac{\beta^{2} gH}{4\pi |l|}
  \right]
\\[4pt]
&\;
- \frac{(gH)^{3/2}}{4\pi^{3/2} \beta}
  \sum_{k=2}^{\infty}
  \frac{2^{2k} B_{2k}}{(2k)!}
  \Gamma\!\left( 2k - \frac{3}{2} \right)
  \sum_{l}
  \frac{(\beta^{2} gH)^{2k - 3/2}}
       {\big[ \beta^{2} gH + ( \beta \Omega_I - 2\pi l )^{2} \big]^{2k - 3/2}} ,
\end{aligned}
\end{equation}
where $V_{\Omega_I}$ denotes the contribution that depends solely on $\beta\Omega_I$,
\begin{equation}\label{eq:VOmegaI}
V_{\Omega_I}
=
-\frac{6}{\pi^{2}\beta^{4}}
\left(
\frac{\pi^{4}}{90}
- \frac{\pi^{2}\beta^{2}\Omega_I^{2}}{12}
+ \frac{\pi\beta^{3}\Omega_I^{3}}{12}
- \frac{\beta^{4}\Omega_I^{4}}{48}
\right).
\end{equation}
The notation $\sum_{l}^{\prime}$ means that the $1/|l|$ term is excluded when $l=0$. The calculation leading to Eq.~\eqref{eq:VhighT} closely parallels that of Ref.~\cite{Meisinger:2002ji}.
The result can be obtained straightforwardly by replacing the Polyakov-loop phase $\phi$ in that reference with $\beta\Omega_I$.  
For this reason, we do not repeat the detailed derivation here. It should be emphasized that, in writing $V_{\Omega_I}$, we have included the contribution from the neutral gauge fields, which are likewise affected by imaginary rotation. 
As a result, the total contribution carries an overall coefficient of $6$ in Eq.~\eqref{eq:VOmegaI}.

In the small-$\Omega_I$ regime, the imaginary angular velocity and chromomagnetic condensate satisfy $\beta\Omega_I < \beta\sqrt{gH} \ll 2\pi$, which is clearly supported by the numerical analysis in Sec.~\ref{sec:numerical} and justifies treating $\Omega_I$ and $gH$ as small expansion variables. Accordingly, the real part of the effective potential can be expanded by retaining terms up to $O(\Omega_I^2)$ and $O(g^2 H^2)$, yielding
\begin{equation}\label{eq:VRexpansion}
\begin{aligned}
V_R
=&\;
\frac{6}{\pi^{2}\beta^{4}}
\left(
\frac{\pi^{4}}{90}
- \frac{\pi^{2}\beta^{2}\Omega_I^{2}}{12}
\right)
\\[4pt]
&\;
+ \Bigg[\frac{1}{2g^2}
-\frac{11}{24\pi^{2}}
\left(
\ln\frac{\beta\mu}{4\pi}
- \gamma
\right)
+ \frac{7 + 4C_{1}}{32\pi^{2}}
- \frac{11}{96\pi^{4}}\,
\zeta(3)\,\beta^{2}\Omega_I^{2}
\Bigg]
(gH)^{2}
\\[4pt]
&\;
- \frac{C_{2}}{2\pi\beta}\,(gH)^{3/2}
- \left(
\frac{11}{24\pi\beta^{2}}
- \frac{C_{3}}{2\pi\beta^{2}}
\right)
\beta^{2}\Omega_I^{2}\, gH .
\end{aligned}
\end{equation}
Here the coefficients $C_1$ and $C_2$ have been determined previously in Refs.~\cite{Meisinger:2002ji,Ninomiya:1981eq,Persson:1996zy} and are given by
\begin{equation}
\begin{aligned}
C_{1}
&=
\sum_{k=2}^{\infty}
\frac{2^{2k} B_{2k}}{(2k)!}
\int_{0}^{\infty} \mathrm{d}t\;
t^{\,2k-3}\, \mathrm{e}^{-t}
\simeq -0.01646,
\\[4pt]
C_{2}
&=
\frac{5}{6}
+
\frac{1}{2\sqrt{\pi}}
\sum_{k=2}^{\infty}
\frac{2^{2k} B_{2k}}{(2k)!}
\Gamma\!\left( 2k - \frac{3}{2} \right)
\simeq 0.82778 .
\end{aligned}
\end{equation}
The coefficient $C_3$ is a new result of the present work.
It originates from the imaginary-rotation contribution to the effective potential and is given by
\begin{eqnarray}
	C_3&=&\frac{1}{2\pi^{\frac{1}{2}}}\sum_{k=2}^\infty\frac{2^{2k}B_{2k}}{(2k)!}\Gamma(2k-1/2)\approx -0.0110613.
\end{eqnarray}
Minimizing the high-temperature expansion of $V_R$ in Eq.~\eqref{eq:VRexpansion}, we find that $gH$ increases with $\Omega_I$ in the small-$\Omega_I$ regime, as shown in Fig.~\ref{fig:small_omega}.
For completeness, we briefly comment on the imaginary part of the effective potential.
In the small-$\Omega_I$ regime, the effective potential always has a nonvanishing imaginary part, which is given by
\begin{equation}
	V_I=-\frac{gH}{2\pi\beta^2}\sqrt{\beta^2gH-\beta^2\Omega_I^2}.
\end{equation}

The coefficient of the $(gH)^2$ term in Eq.~\eqref{eq:VRexpansion}
can be identified with $1/(2g_{\mathrm{eff}}^{2})$,
where $g_{\mathrm{eff}}$ denotes an effective coupling constant~\cite{Chodos:1988be}.
This leads to
\begin{equation}
g_{\mathrm{eff}}^{-2}(T,\Omega_I)
=
g(\mu)^{-2}+\Bigg[
-\frac{11}{12\pi^{2}}
\left(
\ln\frac{\beta\mu}{4\pi}
- \gamma
\right)
+ \frac{7 + 4C_{1}}{16\pi^{2}}
- \frac{11}{48\pi^{4}}\,
\zeta(3)\,\beta^{2}\Omega_I^{2}
\Bigg].
\end{equation}
As $\beta\mu<1$, the first two terms in the square brackets are always positive and thus the effective coupling $g_{\mathrm{eff}}$ increases with imaginary rotation,
signaling an enhancement of infrared interactions.
As a consequence, the system tends to move away from the perturbatively deconfined regime.
This qualitative tendency is consistent with the results obtained in the previous section
as well as with earlier studies~\cite{Chen:2022smf,Chen:2024tkr},
which demonstrated that imaginary rotation can induce a confined phase even at high temperature.
From this perspective, the enhancement of $g_{\mathrm{eff}}$ suggests that the critical temperature is expected to increase with $\Omega_I$. 

Another quantity that can be inferred from Eq.~\eqref{eq:VRexpansion}
is the moment of inertia, defined through the curvature of $V_R$
with respect to the imaginary angular velocity.
We obtain
\begin{equation}\label{eq:moi}
I
=
-\frac{\partial^{2} V_{R}}{\partial (i\Omega_I)^{2}}
=
\frac{1}{\beta^{2}}
\left[
1
- \frac{11 - 12 C_{3}}{12\pi}\,\beta^{2} gH
- \frac{11}{48\pi^{4}}\,\zeta(3)\,(\beta^{2} gH)^{2}
\right].
\end{equation}
An interesting observation is that the contribution from the chromomagnetic
condensate is negative, implying a reduction of the moment of inertia.
Physically, this behavior indicates that chromomagnetic correlations
tend to oppose rotational polarization, thereby suppressing the rotational
response of the system.
A similar mechanism has been suggested in Ref.~\cite{Braguta:2023tqz,Braguta:2023yjn}
to explain the emergence of a negative moment of inertia in rotating
lattice QCD simulations, where this phenomenon was referred to as the
\textit{negative Barnett effect}. 
We emphasize that, within the present framework, the chromomagnetic-field
contribution is subleading in the high-temperature expansion.
Consequently, it is not large enough to flip the sign of the moment of inertia,
so that \(I\) remains positive in the parameter region considered here.
For the same reason, this contribution does not alter the overall tendency of
the critical temperature found in previous studies with only a Polyakov-loop
background~\cite{Chen:2022smf,Chen:2024tkr}. Nevertheless, the negative
contribution found here may provide a hint of the mechanism behind the negative
moment of inertia observed in rotating lattice QCD simulations.

\section{Conclusions and discussions}\label{sec:discussion}

In this work, we have investigated the finite-temperature SU(2) Savvidy model under an imaginary angular velocity, incorporating both a chromomagnetic condensate and a Polyakov-loop background.
We find that, at high temperature, a finite imaginary angular velocity induces a nonzero Polyakov-loop phase and enhances the chromomagnetic condensate.
This behavior is consistent with earlier observations that imaginary rotation tends to favor confinement configurations in Refs.~\cite{Chen:2022smf,Chen:2024tkr}.

Most importantly, by performing a high-temperature expansion, we have extracted the explicit dependence of the effective coupling constant on the imaginary angular velocity.
Our results show that the effective coupling increases with imaginary rotation, signaling an enhancement of infrared interactions.
This finding offers a possible interpretation of the confinement-enhancing tendency mentioned above.
In addition, the curvature of the effective potential with respect to the imaginary angular velocity reveals a negative contribution to the moment of inertia arising from the chromomagnetic condensate.
This result suggests that chromomagnetic fields may play an important role in understanding recent lattice observations of rotating gluonic matter, where a negative moment of inertia has been reported.

Several comments are in order.
First, in this work we do not attempt to eliminate the Nielsen--Olesen instability, which has been extensively discussed in the literature.
Various approaches have been proposed to address this issue~\cite{Bordag:2022ysn,Kondo:2006ih,Parthasarathy:2006is,Vercauteren:2007gx}, and we expect that such methods can also be applied in the present setup with imaginary rotation.
Nevertheless, our primary interest here lies in the structure of the real part of the effective potential and the physical insights that can be gained from the Savvidy model under imaginary rotation.

Second, for real rotation the system is expected to exhibit additional instabilities originating from Landau quantization.
These instabilities may persist even in the presence of boundary conditions~\cite{Cao:2020pmm}, in contrast to the case of pure rotation without background fields, where rotational effects are absent at zero temperature once appropriate boundary conditions are imposed~\cite{Ebihara:2016fwa}.
A detailed analysis of such instabilities lies beyond the scope of the present work and deserves further investigation.
Alternatively, one may formally perform an analytic continuation of the imaginary angular velocity to real values at the final stage of the calculation.
Under this procedure, the moment of inertia takes the same form as that given in Eq.~\eqref{eq:moi}, and in particular the negative contribution arising from the chromomagnetic condensate remains unchanged.
At the same time, the effective coupling constant decreases with increasing real angular velocity, indicating that real rotation tends to favor deconfinement.
This behavior is consistent with previous model studies, although it appears to differ from results obtained from analytically continued lattice simulations.

We also note that this conclusion differs from that of Ref.~\cite{Jiang:2021izj}, where the effective coupling was extracted from the vacuum energy by summing over zero-temperature modes and was found to increase with real rotation.
In our formulation, the vacuum contribution is obtained by taking the zero-temperature limit of the thermodynamic potential.
However, due to the instabilities arising from the combined presence of a chromomagnetic field and rotation discussed above, this limit must be treated with care and may lead to a different result.

Finally, several limitations of the present study should be emphasized and are left for future investigations.
(i) Our analysis is restricted to the SU(2) gauge group.
The formalism developed here can be straightforwardly extended to the SU(3) case, and the qualitative features identified in this work are expected to persist.
The inclusion of quark degrees of freedom is also feasible within the same framework, although such extensions introduce additional dynamical variables and significantly increase the numerical complexity.
(ii) We have focused on the system center and assumed spatially homogeneous background fields in its vicinity.
Far from the center, orbital angular momentum naturally induces spatial inhomogeneity, and the full radial dependence of the effective potential may reveal additional nontrivial structures, even in the absence of a chromomagnetic condensate~\cite{Chen:2024tkr}.
(iii) We have assumed that the chromomagnetic condensate is (anti)parallel to the angular velocity, motivated by the expectation that rotation may polarize gluonic degrees of freedom.
For nonparallel configurations, however, obtaining analytic results becomes considerably more difficult and requires further study.
(iv) Our analysis is performed at the one-loop level.
While technically challenging, it would be interesting to extend the present study to two-loop order~\cite{Bordag:2020xzo,Bordag:2022ysn}, especially since gluon interaction vertices acquire explicit dependence on the angular velocity in a rotating background.

Despite these limitations, the present work highlights the importance of the magnetic component of gauge fields in rotating gluonic matter.
We hope that this study will stimulate further analytical and lattice investigations of rotating gauge theories and contribute to a deeper understanding of the discrepancies between theoretical approaches and lattice simulations.

\section*{acknowledge}
The authors thank Kenji Fukushima, Mei Huang, Guo-Liang Ma, and Kun Xu for useful suggestions and  discussions. This work is supported by the Natural Science Foundation of Shanghai (Grant No. 23JC1400200), the National Natural Science Foundation of China (Grants No. 12225502 and No. 12147101), and the National Key Research and Development Program of China (Grant No. 2022YFA1604900). Recently, we learned that Ref.~\cite{Zhang:2026ctc} addresses a similar topic, which was posted on arXiv on the same day.

\bibliography{ref}
\end{document}